# Quantum Phase Transitions in $Ba_{(1-x)}Ca_xFe_{12}O_{19}$ ($0 \leq x \leq 0.10$)

Keshav Kumar and Dhananjai Pandey*
School of Materials Science and Technology, Indian Institute of Technology (Banaras Hindu University), Varanasi-221005, India

**ABSTRACT**

The ground state of $BaFe_{12}O_{19}$ (BFO) is controversial as three different quantum states, namely quantum paraelectric, frustrated antiferroelectric and quantum electric dipole liquid (QEDL), have been proposed. We have investigated the quantum critical behavior of BFO as a function of chemical pressure (a non-thermal variable) generated by smaller isovalent ion $Ca^{2+}$ at the $Ba^{2+}$ site. Analysis of synchrotron x-ray diffraction data confirms that $Ca^{2+}$ substitution generates positive chemical pressure. Our dielectric measurements reveal that $Ca^{2+}$ substitution drives BFO away from its quantum critical point (QCP) and stabilizes a quantum electric dipolar glass state whose dielectric peak temperature ($T_c$) increases with increasing $Ca^{2+}$ content as $T_c \sim (x-x_c)^{1/2}$, a canonical signature of quantum phase transitions. Our dielectric measurements reveal that pure BFO is slightly away from its QCP with a $T_c$ of 2.91 K. Specific heat measurements reveal excess specific heat of non-Debye and non-magnetic origin with linear temperature dependence below $T_c$ which could be due to QEDL state of BFO.

## 1. Introduction

Classical phase transitions (CPT), like boiling of water, are ubiquitous in nature and affect our day-to-day life. They occur at a finite transition temperatures ($T_c$) as a result of a competition between the energy of the low temperature phase and the entropy fluctuations with an energy scale of $k_B T$. Quantum phase transitions (QPT), on the other hand, occur at the quantum critical point (QCP) $T_c = 0$ K (absolute zero) as a result of the competition between the energy of the system and the quantum fluctuations of energy scale ℏω, where ω is the zero point vibrational frequency of quantum oscillator. Although the QCP ideally occurs at absolute zero temperature which is not experimentally realizable, its remarkable impact on several exotic phenomena occurring at finite temperatures (e.g. high temperature superconductivity, metal insulator transition, integer and fractional Hall effects) makes it an area of vigorous research in condensed matter and materials science [1–7]. The QCP has been located in a variety of strongly correlated [8–12] and other systems ($KH_2PO_4$) [13] by gradually tuning the $T_c$ close to 0 K through non-thermal variables, like composition (x), pressure (p), electric (E) or magnetic (H) field [13–17]. The finger prints of the QPT persist above the QCP also in the so-called "quantum critical" regime also where quantum fluctuations dominate over the thermal fluctuations (i.e. $ℏω > k_BT$) and give rise for example to non-classical exponents for the dependence of $T_c$ on the non-thermal control parameters. Interestingly, Nature has provided us with a family of materials called quantum paraelectrics (QPE), like $SrTiO_3$ [18], $KTaO_3$ [19], $CaTiO_3$ [20], $TiO_2$ [21] and $EuTiO_3$ [22], which already possess a QCP and one can study the effect of quantum fluctuations in the quantum critical regime by driving the system away from the QCP on application of non-thermal variables [16,23].

Very recently, in a series of publications [24,25] it has been argued that M type hexaferrites, especially the $BaFe_{12}O_{19}$ (BFO), are also QPE. First principles calculations on BFO have revealed 1D chains of electric dipoles arranged on a triangular lattice with ferroelectric (FE) and antiferroelectric (AFE) interactions along the c-axis and in the *ab*- plane, respectively, arising out of two unstable Γ point optical phonon modes [26] ( see section S1 and Fig.S1 of the supplemental material for details of crystal structure of BFO). On account of the AFE interactions between the neighbouring dipoles on the triangular lattice, BFO is geometrically frustrated [26,27]. The concept of frustration in the field of ferroelectrics and relaxors is not new but the frustration in BFO is unique because of its geometrical nature as the previously reported systems were based on competing AFE and FE interactions [28–31]. The reports on quantum saturation of $\varepsilon(T)$ in BFO below T ~ 10 K (a canonical signature for the quantum paraelectricity) have become questionable because of the observation of an upturn in the $1/\varepsilon'(T)$ versus T plot around 4 K with a $1/T^3$ like dependence of $\varepsilon'(T)$ in agreement with the theory of uniaxial quantum paraelectrics [32]. First principles DFT calculations in conjunction with MC simulations using a simple dipole-dipole interaction model also suggest the possibility of a phase transition at $T_c$ ~ 3K to a geometrically frustrated antiferroelectric (AFE) state [26]. The complex interplay of geometrical frustration and quantum fluctuations in BFO offers tremendous potential for the discovery of new exotic states of matter with phenomenological similarities with their magnetic counterparts [2,5,33–36] notwithstanding the intrinsic differences in the microscopic origin and dynamics of electrical and magnetic dipoles [27]. In fact, it has already been proposed that the exotic quantum spin liquid like state involving electric dipoles may exist in BFO at very low temperatures [27].

In the present work, we have investigated the effect of a non-thermal variable, namely chemical pressure and local electric field, generated through substitution of $Ba^{2+}$ (r = 1.49Å) with a much smaller isovalent ion $Ca^{2+}$ (r = 0.99Å), on the quantum critical behavior of BFO using dielectric and specific heat measurements from 1.66 to 100 K and 1.8 to 300 K, respectively. We observe a cusp in the dielectric permittivity of BFO at $T_c \sim 2.91$ K with negative Curie-Weiss ($T_{CW}$) temperature expected for antiferroelectric (AFE) correlations, in excellent agreement with the theoretical predictions [26]. By analyzing the specific heat data with and without magnetic field, we have found evidence for excess specific heat of non-Debye and non-magnetic origin which is approximately linear in T below $T_c = 2.91$ K supporting the possibility of a QEDL state at low temperatures in agreement with the earlier report based on thermal conductivity measurements [27]. The $T_c$ of BFO is shown to increase with increasing $Ca^{2+}$ content (x) in the $(x-x_c)^{1/2}$ manner, characteristic of phase transitions in the quantum critical regime [16,37–39] in marked contrast to the $T_c \sim (x-x_c)$ type dependence expected for the classical regime [16,39]. Our dielectric results reveal that $Ba_{(1-x)}Ca_xFe_{12}O_{19}$ (BCFO-x) does not exhibit long range ordered AFE ground state, but may correspond to QEDL like state for pure BFO (x=0), and quantum dipole glass state with frustrated AFE interactions for higher $Ca^{2+}$ content (x ≥ 0.05).

## 2. Experimental Details and Analysis:

The Barium Hexaferrite ($BaFe_{12}O_{19}$) and $Ca^{2+}$ doped $BaFe_{12}O_{19}$ were synthesized by the solid state thermo-chemical reaction route using analytical reagent grade chemicals: $BaCO_3$ (≥ 99.0% assay, Sigma Aldrich), $Fe_2O_3$ (≥ 99.0% assay, Sigma Aldrich) and $CaCO_3$ (≥99.5% assay, Alfa Aesar). Stoichiometric mixture of chemicals were mixed properly using an agate mortar and pestle for 3 hours followed by ball milling in zirconia jar and balls which acted as the grinding medium. Acetone was used as a milling medium. Mixing was done for 12 hours. The sample was dried at room temperature after mixing. Calcinations were carried out in alumina crucibles. Pure $BaFe_{12}O_{19}$ was synthesized by calcination at $1100^0$C for 6 hour. The $Ca^{2+}$ doped samples were calcined at $1250^0$C for 6 hour. The calcined powders were further crushed using mortar and pestle. Then a few drops of 2% Polyvinyl Alcohol (PVA) were added to it as a binder. Pellets were made using cylindrical steel die and uniaxial hydraulic press machine. Pellets were kept at $600^0$C for 10 hour to remove the PVA. Pure $BaFe_{12}O_{19}$ sample was sintered at $1200^0$C for 6 hour whereas the $Ca^{2+}$ doped samples were sintered at $1275^0$C for 1 hour. Room temperature SXRD data were collected at a wavelength of 0.207150Å at P02.1 beamline in Petra III, Hamburg, Germany. For the synchrotron x-ray diffraction (SXRD) measurements, the powder used was obtained from sintered pellets which were crushed into fine powders and then annealed at $600^0$C for 10 hours to remove the stresses introduced during crushing. Rietveld refinement was carried out using Fullprof Suite [50].

For dielectric measurements, the top and bottom surfaces of the sintered pellets were mildly polished using diamond paste. After polishing, the pellets were kept in isopropyl alcohol to remove moisture and then electroded using fired-on (500 C for 2 minutes) silver paste . Low temperature dielectric permittivity of $BaFe_{12}O_{19}$ was measured for all the samples using a fully computer controlled measuring system involving Novo Control Alpha-A High Frequency Analyzer and a cryogen free measurement system (CFMS). First the measurements were carried out in the range 1.66 K to 100 K at a heating rate 0.3 K per minute at a fixed frequency of 300 kHz. This was followed by dielectric measurements over the temperature range 1.66 K to 25 K at multiple frequencies in the range 10 kHz to 400 kHz.

For heat capacity measurements, a small piece of sintered pellet was used after annealing at $600^0C$ for 10 hours. The heat capacity measurement of pure $BaFe_{12}O_{19}$ was carried out using a Physical Properties Measurement System DynaCool (Quantum Design).

**2.1. Evidence for chemical pressure generated by $Ca^{2+}$ substitution**

The room temperature crystal structure of BFO is hexagonal in the space group $P6_3/mmc$. Since $Ca^{2+}$ ion is smaller in size than the $Ba^{2+}$, it is expected to generate positive chemical pressure and the unit cell volume should accordingly decrease with increasing $Ca^{2+}$ content (x). To confirm this, we carried out Rietveld refinement using SXRD data for various composition of BCFO-x. The details of refinement are given in the section S1 of the supplemental material. We have obtained excellent fits between the observed and calculated profiles after Rietveld refinement for all the compositions. This is illustrated in Fig.1 for x=0 (i.e. BFO) and x=0.10. Our Rietveld refinements confirm that the structure of BCFO-x (x > 0) remains identical to BFO (x=0), i.e. hexagonal in the $P6_3/mmc$ space group. Figs 2 (a) and 2 (b) depict the variation of lattice parameters (a, c) and unit cell volume (V) as a function of $Ca^{2+}$ content (x), respectively. The fact that the unit cell volume decreases with increasing x confirms that $Ca^{2+}$ substitution generates positive chemical pressure in the BFO matrix. In addition to generating chemical pressure, $Ca^{2+}$ substitution may also lead to creation of local electric dipoles [40] because the smaller ions like $Ca^{2+}$ have a tendency to go off-centre with respect to the $Ba^{2+}$ site in the centre of the $AO_{12}$ polyhedra, as is well known for $Ca^{2+}$ doped $SrTiO_3$ (SCT) [16,41] or $Li^{1+}$ doped $KTaO_3$ (KTL) [19,40,42]. We also calculated the bond lengths in the ab-plane (Fe2-O3) and along the c-axis (Fe2-O1) using Fullprof Suite [50] from the refined positional coordinates. Figs 2 (c) depicts their variation with x. As expected, these bond lengths also decrease with increasing $Ca^{2+}$ content. The refined structural parameters and bond lengths for all the compositions are listed in Table S1 of the supplemental material.

**2.2. Effect of $Ca^{2+}$ substitution on quantum critical behaviour of BFO**

To understand the effect of chemical pressure and local electric field due to off-centred $Ca^{2+}$ ions on the quantum critical behaviour of BFO, we show in Fig. 3 the temperature dependence of the real $\varepsilon'(T)$ and imaginary $\varepsilon''(T)$ parts of the dielectric permittivity of BCFO-x in the temperature range 1.66 to 100 K measured at 300 kHz. It is evident from Fig. 3(a) that pure BFO shows a smeared dielectric response with a critical temperature $T_c \sim 2.91$ K which is in excellent agreement with the theoretical predictions of a transition at $T_c \sim 3$ K [26]. $Ca^{2+}$ substitution enhances $T_c$ and also makes the dielectric anomaly more prominent by increasing the peak height at the critical temperature. The increase in $T_c$ seems to be primarily due to an increase in the strength of the dipole-dipole interactions both in the ab plane as well as along the c axis as a result of the contraction of the bond lengths of the $FeO_5$ trigonal bi-pyramids (TBPs) caused by chemical pressure generated by $Ca^{2+}$ (see Fig.2 (c)). This is consistent with the theoretical predictions also according to which external compressive stresses should in general enhance $T_c$ [26]. The $T_c$ in other doped quantum paraelectric systems, like SCT [16] and KTN [19,42], is known to follow $(x-x_c)^{1/2}$ type composition dependence in the quantum critical regime [16,38,39,42] in marked contrast to $(x-x_c)$ type dependence for the classical phase transitions [39,42]. The variation of $T_c$ of BCFO-x as a function of $Ca^{2+}$ content (x) also follows $(x-x_c)^{1/2}$ type dependence as can be seen from Fig.4. The continuous line in this figure through the data points is the least squares fit to $(x-x_c)^{1/2}$ type dependence of $T_c$.

Interestingly, the extrapolation of the $T_c \sim (x-x_c)^{1/2}$ plot for BCFO-x to x=0 gives a value of $T_c \sim 2.85$ K for undoped BFO in perfect agreement with the experimentally observed peak temperature $T_c= 2.91$ K shown in the inset of Fig.3(a) and the theoretically predicted $T_c \sim 3$ K [26]. Further, $T_c$ approaches 0 K on the negative side of the composition axis suggesting that negative pressure is required to locate the quantum critical point (QCP) of BFO. This is similar to the situation reported in some heavy fermion systems undergoing quantum phase transitions where QCP lies on the negative pressure axis of the phase diagram [10,43]. Such a negative pressure in BFO can be generated by substitution with a larger isovalent ion like $Pb^{2+}$ and we predict that $Pb^{2+}$ substitution can drive BFO closer to its QCP in contrast to $Ca^{2+}$ which drives it away from the QCP. Our results thus reveal that BFO is very close to QCP, but its true QCP can be realised by applying negative pressure. Before we close this discussion, we would like to add that BCFO-x compositions exhibit another transition around 20 K seen in Fig.3 linked with a magnetic phase transition which is the subject matter of a separate publication.

### 2.3. Evidence for quantum electric dipole glass state in $Ca^{2+}$ substituted BFO

The positive chemical pressure and/or local electric field generated by random site dipoles associated with off-centred cations are known to suppress quantum fluctuations [39] and stabilise quantum ferroelectric/ferrielectric phases in SCT [16,41] and quantum dipole glass in KLT [19,40]. Unlike the SCT and KLT systems, the geometrical frustration in $Ca^{2+}$ doped BFO (e.g. BCFO-x) may lead to three different types of phases below the dielectric peak temperature $T_c$: (1) an LRO state as observed in SCT [16,41] and predicted for BFO also as in Ref [26], (2) a quantum dipole glass state with a characteristic critical dynamics showing ergodicity breaking similar to KLT [40] and (3) quantum electric dipole liquid (QEDL) state as proposed in Ref [27]. The $\varepsilon'$ ($\omega$, T) plots near the peak temperature $T_c$, measured at several frequencies (in the 40 to 400 kHz) at a very slow heating rate of 0.1 K/min, reveal considerable dispersion for all the compositions (see the insets of Fig.3). Further, the temperature $T'_m$ corresponding to the peak in $\varepsilon'$ ($\omega$, T) for x ≥ 0.03 shifts to higher temperatures on increasing the measuring frequencies ($\omega=2\pi f$). Figs. 5(a) and 5(b) depict the $\varepsilon'$ ($\omega$, T) plots for a wider frequency range for x =0 and x=0.05, respectively, on a magnified scale for better clarity. It is evident from these figures that the dielectric peak temperature shifts to higher frequencies for the doped samples only as no such shift is observed for the undoped BFO down to 500 Hz. However, even undoped BFO shows considerable dispersion in the value of $\varepsilon'$ indicating highly degenerate ground state with possible low temperature tunnelling among the various states. Frequency dependent shift of the dielectric peak temperature is known to occur in glassy systems [19, 40]. The non-linear nature of the ln($\tau$) versus 1/T plot shown in Fig.6 for x= 0.05 rules out Arrhenius behaviour of the relaxation time $\tau$. A similar situation holds good for x = 0.07 and 0.10 also. We could model the temperature dependence of $\tau$ of all the composition for x > 0.03 using power law dynamics, commonly used in spin glass systems [44]:

$$\tau = \tau_0 \left(\frac{T_{max}-T_g}{T_g}\right)^{-z\nu} \quad \ldots(1)$$

where $\tau_0$= 1/$\omega$ is the inverse of the attempt frequency, $T_g$ is the critical temperatures at which the slowest polar dynamics diverges signalling ergodicity breaking and $z\nu$ is the critical exponent related to the correlation length. The fit corresponding to the power law behaviour is shown in the inset of Fig.6 for x=0.05. The critical exponent increases from $zv = 0.94 \pm 0.002$ K to $1.89 \pm 0.004$ K on increasing $Ca^{2+}$ content from x=0.05 to 0.10 but remains within the limit of canonical glasses as discussed in the context of the spin glass systems. [44] Similarly, the attempt frequency also increases from 9.9 x $10^6$ to 1.08 x $10^8$ with $Ca^{2+}$ content. Our analysis clearly shows that $Ca^{2+}$ substitution induces a crossover from diffuse transition behaviour of BFO, evidenced by the smeared ε'(T) response for x=0 with absence of frequency dependent shift of $T_c$, to a canonical quantum dipole glass phase. We find that the glass freezing temperature $T_g$ also shows $(x-x_c)^{½}$ type composition dependence (see inset of Fig. 4) with a better fit as compared to that for $T_c$ at 300 kHz shown in the main figure.

## 2.4. Evidence for QEDL phase

The dielectric permittivity of BFO is anisotropic with a value of $ε_c$ ~ 42 and $ε_{ab}$ ~ 18.2 along the c-axis and the ab- plane, respectively [24]. The QPE behaviour (i.e. increase in the dielectric permittivity followed by its approximate saturation below ~10 K) is shown by $ε_c$ only due to softening of the transverse optical phonon mode of $A_{2u}$ symmetry [26,45] imparting BFO a uniaxial character with dipoles aligned along the *c*-axis. The value of dielectric permittivity of BFO in our case is ~ 33.67 at ~ 3K which in effect corresponds to an average over a large number of polycrystalline grains each with $ε_c$ ~ 42 and $ε_{ab}$ ~ 18.2. The $ε_c$ of BFO increases with decreasing temperature and reported to saturate below ~10K whereas the $ε_{ab}$ decreases and saturates to a nearly constant value for T < 50K [24]. This suggests that the temperature dependence of dielectric permittivity ε (T) of sintered polycrystalline BFO below 50K would be essentially due to the temperature dependence of $ε_c$ only and we can analyse the data to examine the critical exponents of undoped and doped BFO in the low temperature range up to 50 K.

The Curie-Weiss temperature $T_{CW}$, obtained from the extrapolation of the linear region in Curie-Weiss plots upto ~40 K (see Fig. 7) is found to be negative for all the compositions in the range $0 \leq x \leq 0.1$ as expected for AFE correlations. The magnitude of $T_{CW}$ increases with increasing $Ca^{2+}$ content (see Fig. 8). Since $T_c$ of BFO is ~ 3K whereas $|T_{CW}|$ is ~ 423 K, the frustration parameter $f = |T_{CW}| / T_c$ is very high (~141) and lies in the range where quantum spin liquid phase has been reported in the magnetic systems [2,46].

One of the criteria used to define a QSL phase in magnetic systems is linear temperature dependence of specific heat at low temperatures [2,36,47,48]. The temperature dependence of the specific heat of BFO from 1.8 K to 300 K is shown in Fig.9. The specific heat plot does not reveal any sharp anomaly around $T_c$ = 2.91 K, expected for a phase transition to a LRO AFE phase or glassy freezing, as can be seen more clearly from the inset (a) of Fig. 9 on a magnified scale. On the contrary, it shows linear temperature dependence in the 1.8 to 2.5 K temperature range below $T_c$ with a knee around 3.5 K. Such a linear temperature dependence of specific heat points towards the QEDL state in close analogy with QSL state [2,36,47,48]. Even after application of magnetic fields up to 6T, we did not see any departure from the linear behavior suggesting that this feature is linked with electric dipoles only and not the magnetic spins of the lattice. The signature of QEDL state is better seen after subtracting the Debye lattice contribution at low temperatures ($T^3$ dependence) in this non-metallic system. Since the underlying magnetic sublattice is predominantly ferrimagnetic with large magnetic moment, one expects a $T^{3/2}$ dependence for the magnetic LRO phase [49]. Taking both aspects together, the specific heat of BFO should exhibit $C_p = \alpha T^3 + \beta T^{3/2}$ type temperature dependence at very low temperatures. This was confirmed by the linearity of the $C_p/T^{3/2}$ vs $T^{3/2}$ plot shown in the inset (b) of Fig. 9 in the temperature range 1.8 to 11.8 K. Using this fit, we subtracted the Debye contribution to obtain the non-Debye part which is shown in inset (c) of Fig. 9. The non-Debye part of specific heat also shows a linear temperature dependence below the dielectric anomaly peak temperature $T_c \sim 3$ K of undoped BFO. The positive curvature at low temperatures reveals non-zero entropy in the ground state characteristic of a geometrically frustrated systems [2,35,48]. The linear temperature dependence of the non-Debye part of the specific heat below ~3 K is not of magnetic origin as it does not show any field dependence upto 6T (see inset (c) of Fig. 9). Magnetic field leads to a slight decrease in the specific heat above 3 K and transforms the step like feature of non-Debye contribution of zero field measurement to a broad peak (see inset (c) of Fig. 9). The linear temperature dependence of specific heat and its field independence points towards a QEDL state below 3 K, although more work is required at much lower temperatures than those used in the present work to confirm this proposition. The excess specific heat of non-Debye and non-magnetic origin clearly suggests presence of low energy excitations. We hope that our results would encourage inelastic neutron scattering experiments with appropriate theoretical models to understand the precise nature of these excitations and their possible role in the QEDL state.

## 3. Summary

To summarize, we have shown that $BaFe_{12}O_{19}$ exhibits a smeared dielectric response due to AFE correlations peaking at $T_c \sim 3$ K with a frustration parameter f ~141. We have also presented results of specific heat measurements at different magnetic fields which suggest the possibility of a quantum electric dipole liquid state in this compound. We have investigated for the first time the quantum critical behavior of $BaFe_{12}O_{19}$ by driving it away from its quantum critical point (QCP) through a non-thermal variable, namely chemical pressure generated by $Ca^{2+}$ substitution. Using Rietveld analysis of SXRD data, we have shown that the unit cell volume decreases with increasing $Ca^{2+}$ content (x) confirming positive chemical pressure generated by $Ca^{2+}$ substitution. Our dielectric measurements reveal that the chemical pressure generated by $Ca^{2+}$ substitution stabilises a quantum electric dipole glass state whose glass transition temperature ($T_g$) follows $(x-x_c)^{1/2}$ type of composition dependence expected for a quantum phase transition. Our results reveal that $BaFe_{12}O_{19}$ is close to its quantum critical point (QCP) which can be reached by applying negative chemical pressure.

**Acknowledgements:** DP acknowledges financial support from Science and Engineering Research Board (SERB) of India through the award of J C Bose National Fellowship. The authors sincerely acknowledge Prof. T.V. Ramakrishnan for fruitful discussions and reading the manuscript. We acknowledge the support from India-DESY project of the Department of Science and Technology, Govt. of India operated through Jawaharlal Nehru Centre for Advanced Scientific Research, Jakkur, India. We thank beam line scientist Dr. Martin Etter at PETRA III for his help.

**Figure Captions**

**Fig.1.** (a) Observed (red circles), calculated (black continuous line) and difference (bottom green line) profiles obtained after Rietveld refinement using P6$_3$/mmc space group for Ba$_{(1-x)}$Ca$_x$Fe$_{12}$O$_{19}$ with (a) x= 0.00 and (b) x=0.10. The vertical bars represent the Bragg peak positions (blue).

**Fig.2.** Variation of (a) lattice parameters a, c, (b) unit cell volume V and (c) bond lengths Fe2-O1 and Fe2-O3 of Ba$_{(1-x)}$Ca$_x$Fe$_{12}$O$_{19}$ in the composition range $0.00 \leq x \geq 0.10$.

**Fig.3.** Variation of the real ($\varepsilon'$) and imaginary ($\varepsilon''$) parts of the dielectric permittivity of Ba$_{(1-x)}$Ca$_x$Fe$_{12}$O$_{19}$ at 300 kHz for different Ca$^{2+}$ concentrations with x = (a) 0.00, (b) 0.03, (c) 0.05, (d) 0.07 and (e) 0.10. Insets show the variation of the real part of the dielectric permittivity at various frequencies (40 kHz (□), 50 kHz (●), 70 kHz (△), 80 kHz (▼), 100 kHz (◇), 150 kHz (◄), 200 kHz (▷) 250 kHz (●) 300 kHz (☆), 400 kHz (♠))

**Fig.4.** Variation of dielectric peak temperature (T$_c$) at 300 kHz of Ba$_{(1-x)}$Ca$_x$Fe$_{12}$O$_{19}$ as a function of Ca$^{2+}$ concentration (x). Inset shows variation of glass transition temperature (T$_g$) as a function of Ca$^{2+}$ concentration (x).

**Fig.5.** Variation of the real ($\varepsilon'$) part of the dielectric permittivity of Ba$_{(1-x)}$Ca$_x$Fe$_{12}$O$_{19}$ for (a) x=0 and (b) x=0.05 measured at various frequencies (500 Hz (■), 1 kHz (○), 10 kHz (●),50 kHz (▲), 70 kHz (▽), 80 kHz (☆) 100 kHz (◆), 150 kHz (△), 200 kHz (◁) 250 kHz (⊞) 300 kHz (▶).

**Fig.6.** Non-Arrhenius behaviour of temperature dependence of relaxation time ($\tau$) shown in ln $\tau$ versus 1/T plot of Ba$_{(1-x)}$Ca$_x$Fe$_{12}$O$_{19}$ for x=0.05. The continuous line shows fit for the power law dynamics $\tau = \tau_0 (\frac{T_{max} - T_g}{T_g})^{-z\nu}$ characteristic of a dipolar glass transition. The inset shows fit for ln$\tau$ vs ln$(\frac{T_{max}-T_g}{T_g})$ plot.

**Fig.7.** Curie-Weiss fit (black solid line) to temperature dependent permittivity (red circles) of Ba$_{(1-x)}$Ca$_x$Fe$_{12}$O$_{19}$ for x = (a) 0.00, (b) 0.03, (c) 0.05 (d) 0.70 and (e) 0.10.

**Fig.8.** Variation of the magnitude of Curie-Weiss temperature |T$_{CW}$| of Ba$_{(1-x)}$Ca$_x$Fe$_{12}$O$_{19}$ with concentration (x).

**Fig.9.** Specific heat of pure BaFe$_{12}$O$_{19}$ as a function of temperature. Insets: (a) specific heat measured at different fields shown on a magnified scale, (b) C$_p$/ T$^{3/2}$ vs T$^{3/2}$ plot where solid line represents the linear fit, (c) non-Debye part of the specific heat at different magnetic fields.

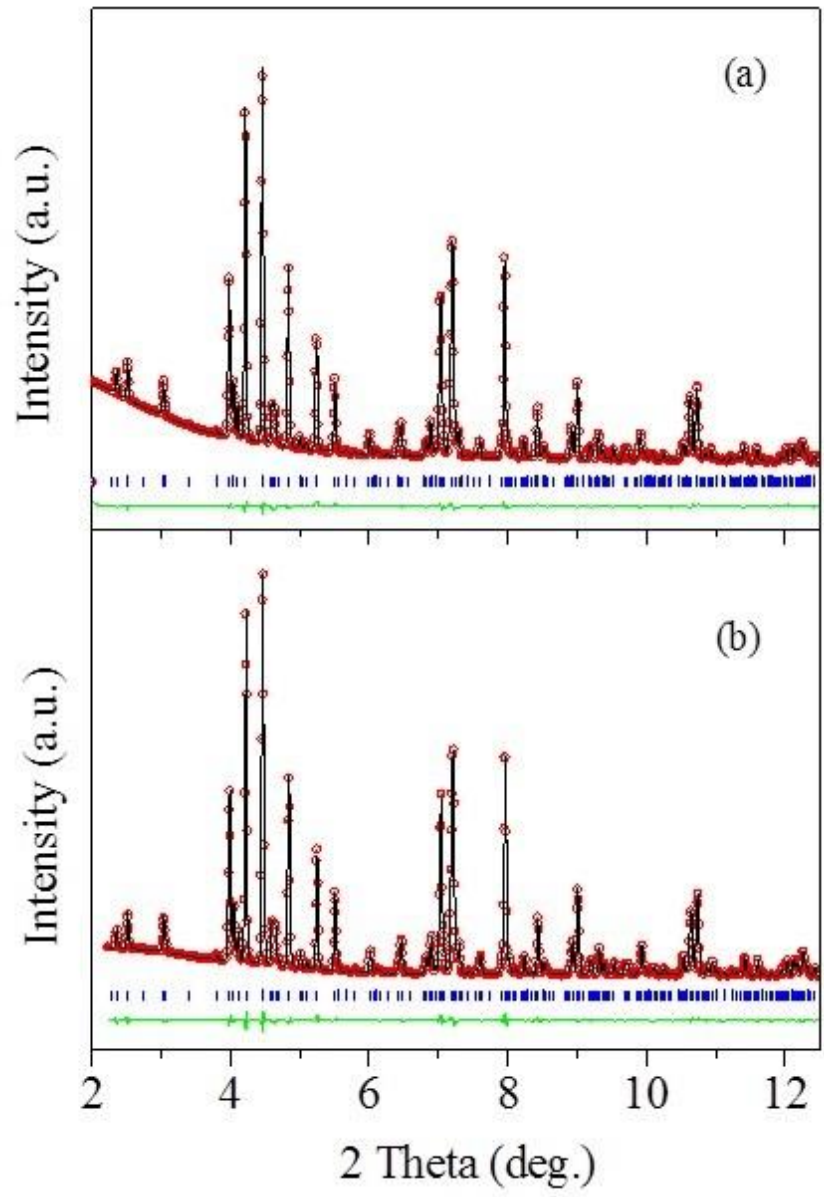

**Figure 1**

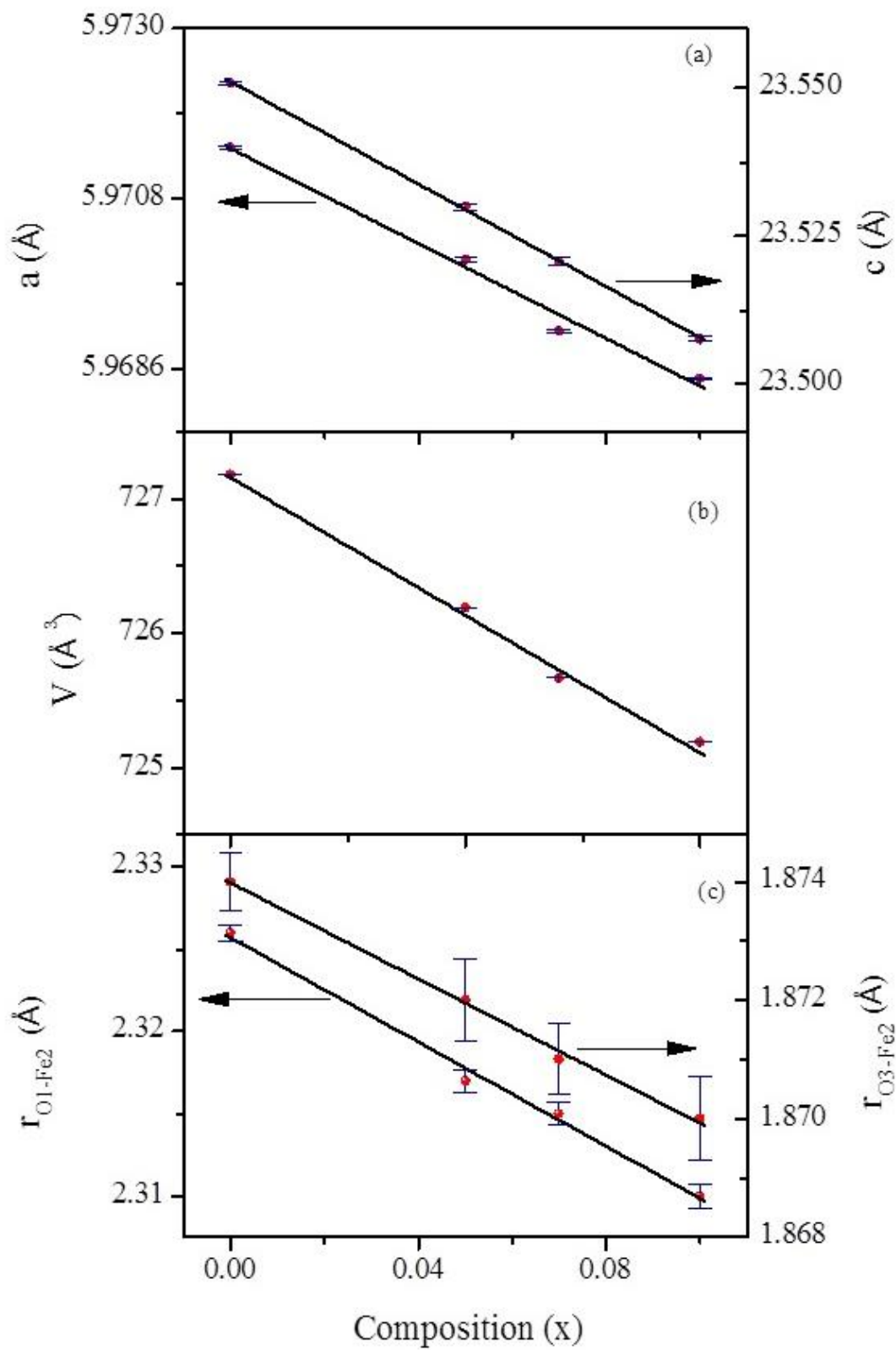

**Figure 2**

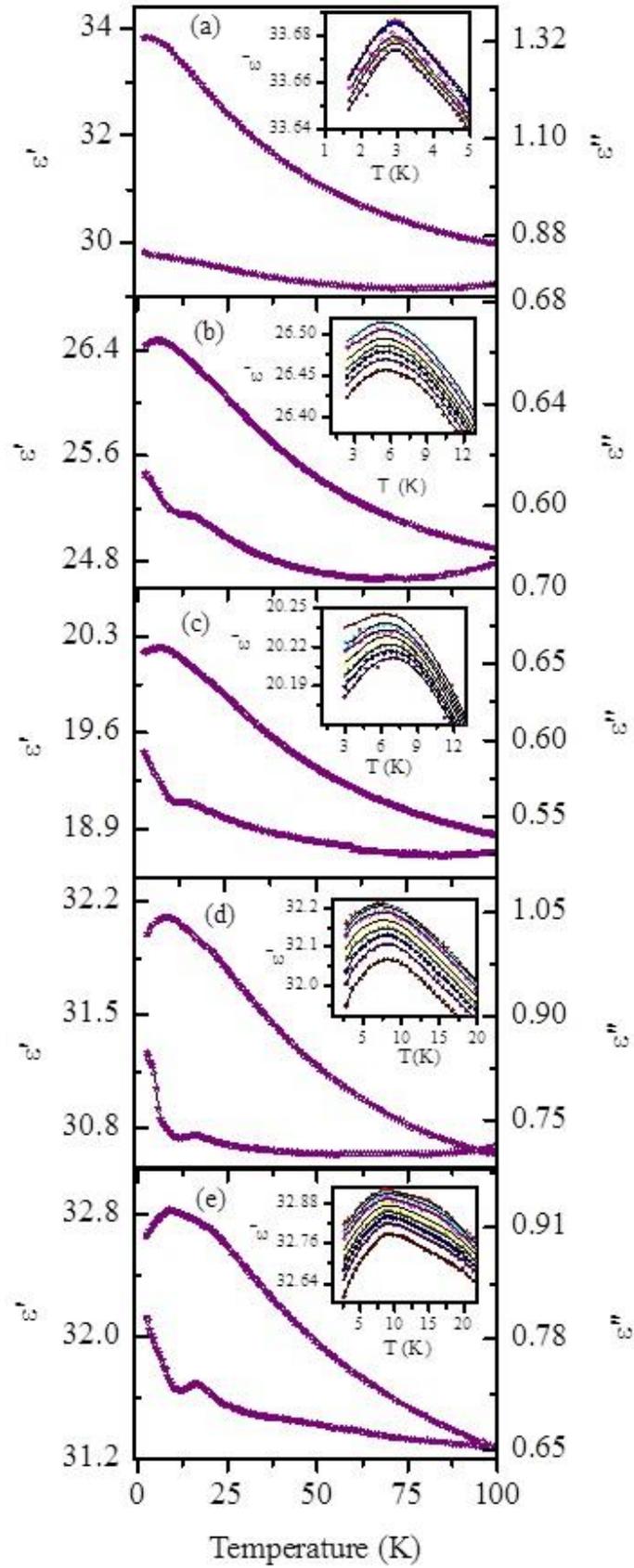

**Figure 3**

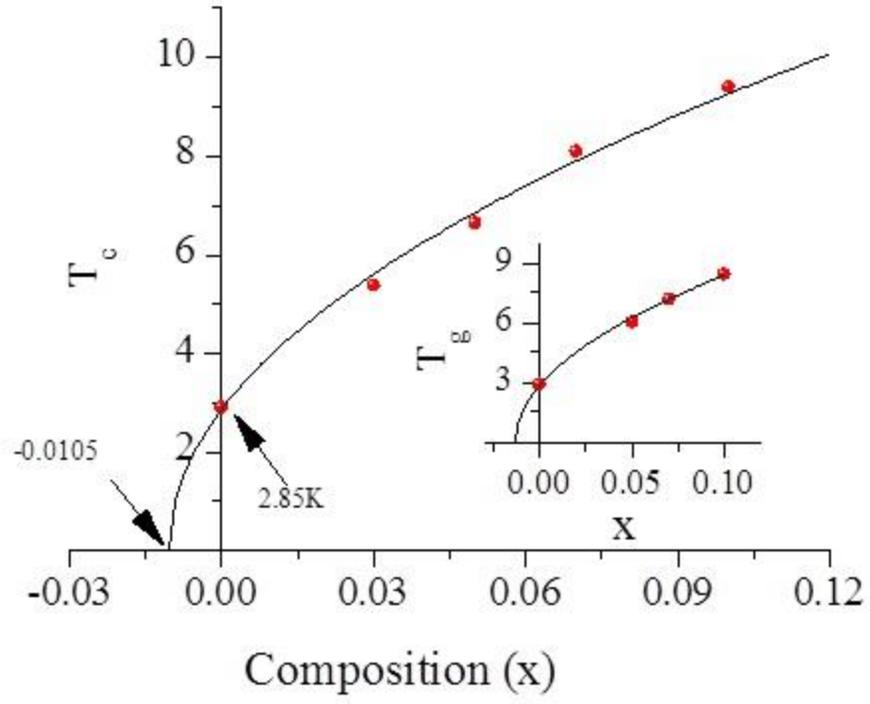

**Figure 4**

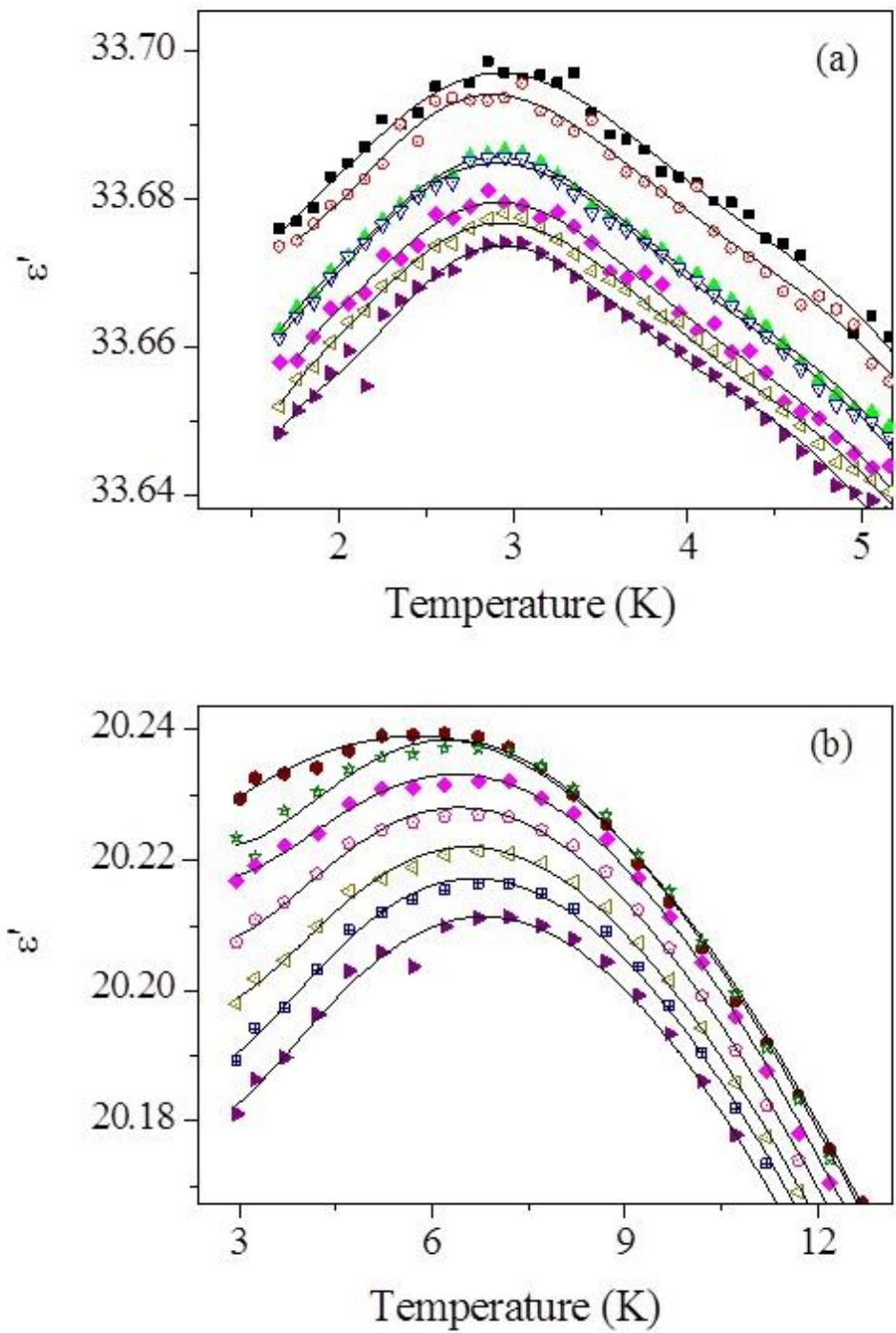

**Figure 5**

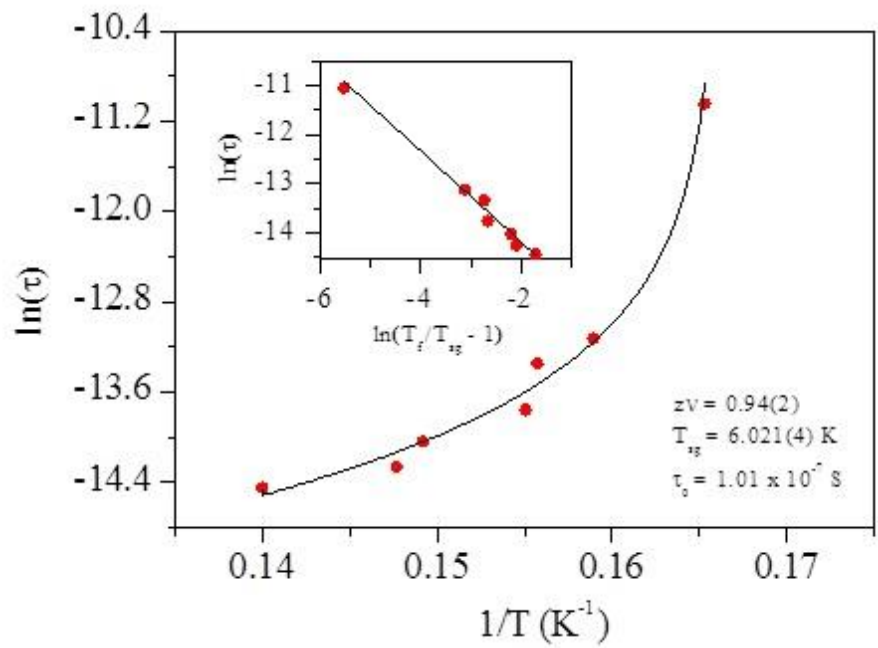

**Figure 6**

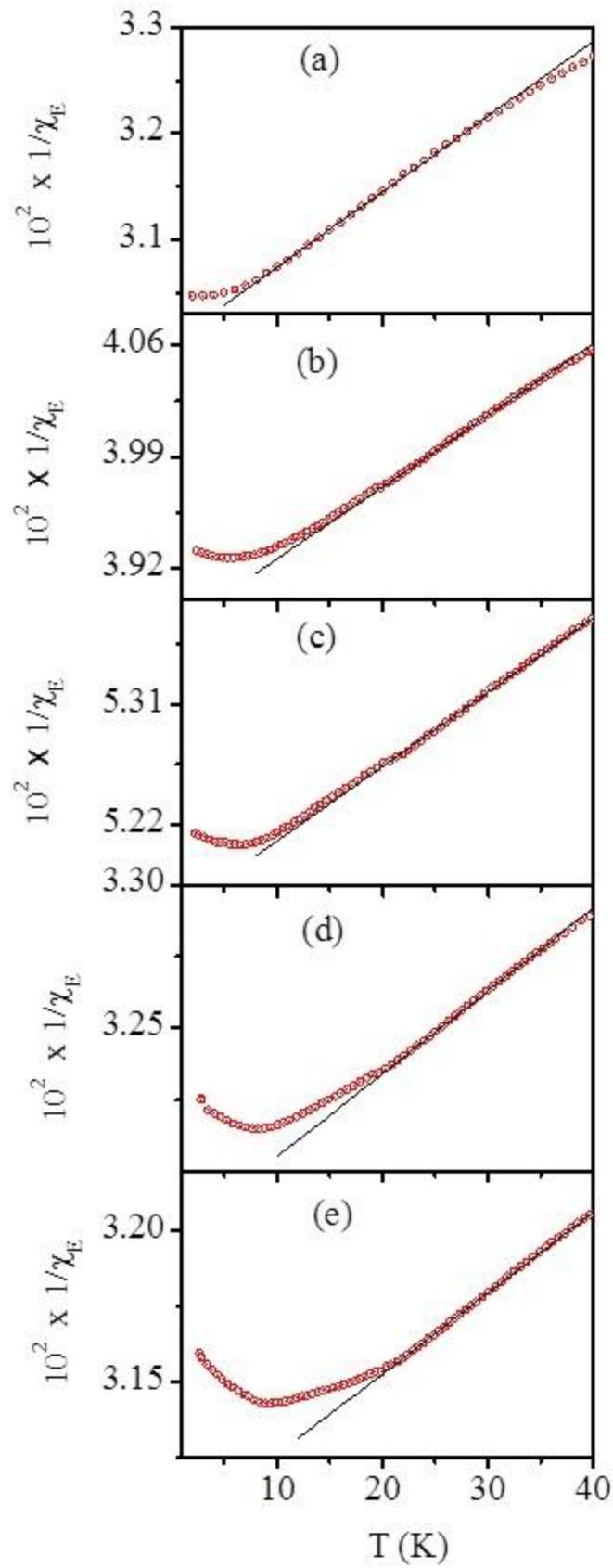

**Figure 7**

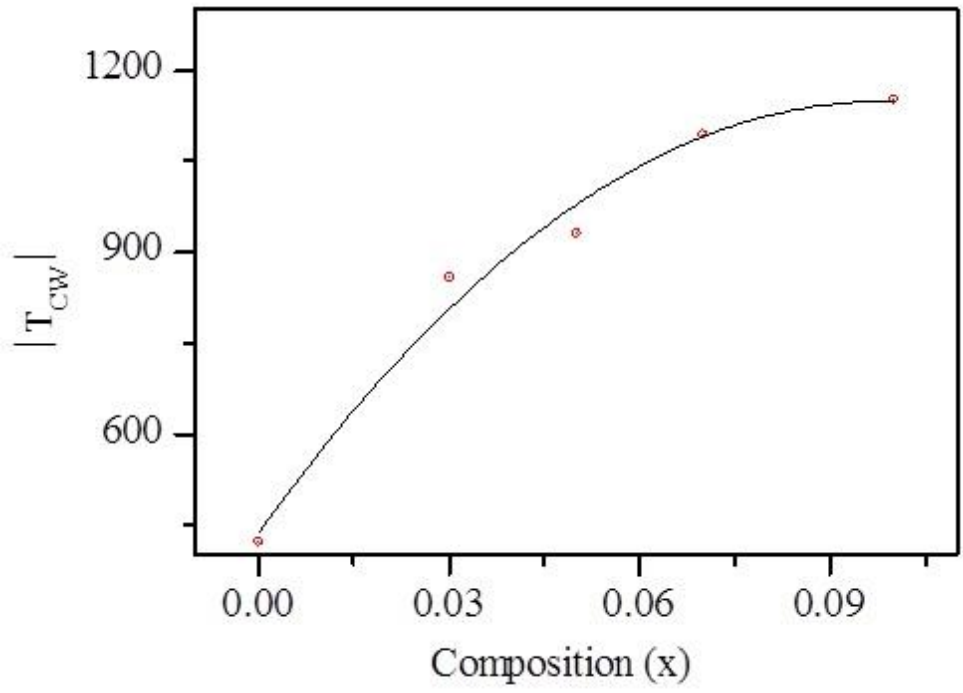

**Figure 8**

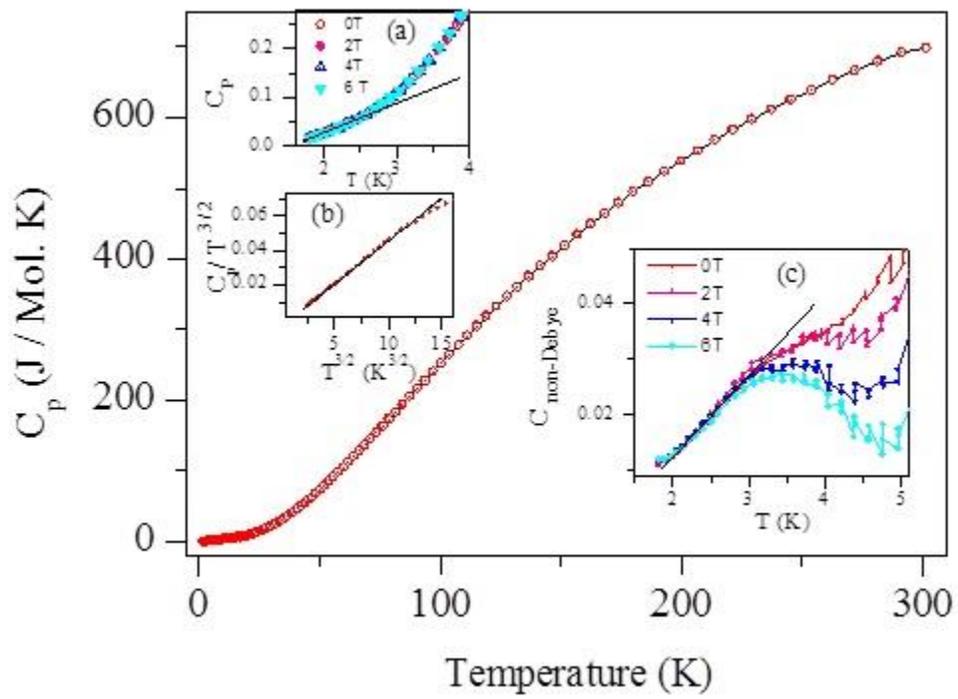

**Figure 9**

**Supplemental material:**

**S1. Structure of BaFe$_{12}$O$_{19}$ and details of Rietveld refinement**

Room temperature crystal structure of BaFe$_{12}$O$_{19}$ is hexagonal in the space group P6$_3$/mmc [S1, S2]. As shown in Fig.S1, the unit cell can be visualized in terms of stacking of S and R blocks in sequence RSR*S* (* means block is rotated 180$^0$ about hexagonal axis w.r.t. initial block). The S block contains two close-packed layers of oxygen in a hexagonal close packed arrangement whereas the R block consists of three close packed layers of oxygen with one oxygen in the middle layer being replaced with Ba. In the hexagonal unit cell, oxygen atoms occupy 4e, 4f, 6h, 12k, 12k Wyckoff positions. The Ba atoms go to the Wyckoff site 2d. Fe atoms occupy five different Wyckoff sites 2a, 4f$_2$, 12k (octahedral sites), 4f$_1$ (tetrahedral site) and 2b (trigonal bi-pyramidal (TBP) site. The Fe spins are ferromagnetically coupled within the "*ab*"- plane but the spins at 2a, 2b, 12k sites are antiparallel to those at the 4f$_1$, 4f$_2$ sites in the *c* direction. Since the number of up (16 spins) and down (8 spins) spins are unequal, BaFe$_{12}$O$_{19}$ shows an Ising type collinear ferrimagneitc ordering of spins [S3] with a large net magnetic moment of 20μ$_B$ per formula unit and a Curie temperature of T$_N$ = 720K. At room temperature, x-ray and neutron as well as first principles studies have revealed that Fe$^{3+}$ ion does not sit at the center of the TBP (at the 2b site) but is displaced away from the a mirror plane to the 4e sites along the *c*-axis [S2, S4, S5].

Rietveld refinement using SXRD data for various composition of BCFO-x were carried out. The positional coordinates of the atoms in the asymmetric unit used in the refinement are : 2/3,1/3,0.25 for Ba at 2d site, 0,0,0 for Fe1 at 2a site, 0,0,0.25 for Fe2 at 2b site, 1/3,2/3,z for Fe3 at 4f1 site, 1/3,2/3,z for Fe4 at 4f2 site, x,y,z for Fe5 at 12k site, 0,0,z for O1 at 4e site, 1/3,2/3,z for O2 at 4f site, x,y,0.25 for O3 at 6h site, x,y,z for O4 at 12k site and x,y,z for O5 at 12k site. Table S1 lists the refined structural parameters for all the compositions.

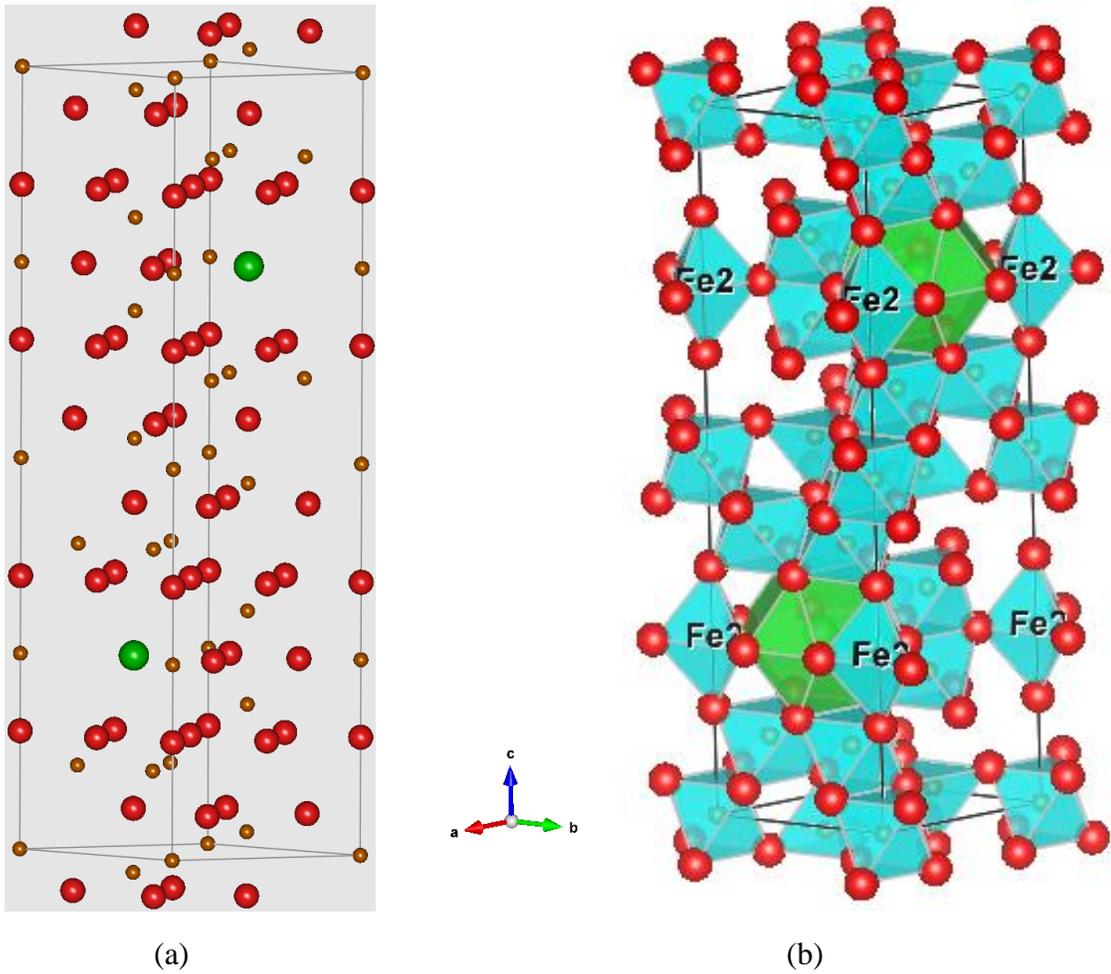

(a)                    (b)

**Fig. S1.** (a) Unit cell of M-Type hexaferrite. (b) Polyhedra of the M-type unit cell

**Table S1:** Atomic positions obtained from the Rietveld refinement for different compositions

| Atoms | x = 0.00 | x = 0.05 | x = 0.07 | x = 0.10 |
|---|---|---|---|---|
| a (Å) | 5.9723(2) | 5.9707(4) | 5.9700(5) | 5.9690(3) |
| c (Å) | 23.540(2) | 23.521(3) | 23.509(3) | 23.501(2) |
| B(Ba/Ca) | 0.5(2) | 0.1(1) | 0.2(2) | 0.3(2) |
| B(Fe1) | 0.3(4) | 0.2(5) | 0.4(5) | 0.4(6) |
| B(Fe2) | 1.1(4) | 1.1(6) | 1.1(5) | 1.4(7) |
| $Z_{Fe3}$ | 0.0272(6) | 0.0273(8) | 0.0272(7) | 0.0272(7) |
| B(Fe3) | 0.3(3) | 0.1(3) | 0.2(3) | 0.3(4) |
| $Z_{Fe4}$ | 0.1904(5) | 0.1904(7) | 0.1904(7) | 0.1904(8) |
| B(Fe4) | 0.4(2) | 0.3(3) | 0.4(3) | 0.4(4) |
| $X_{Fe5}$ | 0.168(2) | 0.168(3) | 0.168(3) | 0.168(3) |
| $Y_{Fe5}$ | 0.337(4) | 0.337(6) | 0.337(5) | 0.337(6) |
| $Z_{Fe5}$ | -0.1082(3) | -0.1083(4) | -0.1083(4) | -0.1084(4) |
| B(Fe5) | 0.5(1) | 0.4(1) | 0.4(2) | 0.4(1) |
| $Z_{O1}$ | 0.151(2) | 0.151(3) | 0.151(2) | 0.151(3) |
| B(O1) | 0.4(1) | 0.0(2) | 0.3(2) | 0.4(2) |
| $Z_{O2}$ | -0.054(2) | -0.054(3) | -0.054(2) | -0.054(3) |
| B(O2) | 0.2(1) | 0.2(2) | 0.1(2) | 0.1(2) |
| $X_{O3}$ | 0.18(1) | 0.18(1) | 0.181(1) | 0.180(1) |
| $Y_{O3}$ | 0.36(2) | 0.36(3) | 0.36(2) | 0.361(3) |
| B(O3) | 0.2(9) | 0.1(1) | 0.2(1) | 0.4(1) |
| $X_{O4}$ | 0.156(7) | 0.157(9) | 0.157(9) | 0.15(1) |
| $Y_{O4}$ | 0.31(1) | 0.31(2) | 0.31(2) | 0.31(2) |
| $Z_{O4}$ | 0.051(1) | 0.052(1) | 0.052(1) | 0.052(2) |
| B(O4) | 0.4(6) | 0.1(8) | 0.2(7) | 0.3(8) |
| $X_{O5}$ | 0.501(9) | 0.50(1) | 0.500(9) | 0.50(1) |
| $Y_{O5}$ | 1.00(2) | 1.00(2) | 1.00(2) | 1.00(2) |
| $Z_{O5}$ | 0.149(1) | 0.149(2) | 0.149(2) | 0.149(2) |
| B(O5) | 0.5(6) | 0.03(9) | 0.3(8) | 0.034(7) |
| $\chi^2$ | 1.13 | 1.12 | 1.47 | 0.785 |
| $R_{wp}$ | 3.15 | 3.90 | 4.06 | 4.19 |
| $R_{exp}$ | 2.97 | 3.68 | 3.79 | 4.73 |
| Fe2-O1 | 2.326(5) | 2.317(7) | 2.315(7) | 2.310(7) |
| Fe2-O3 | 1.874(5) | 1.872(7) | 1.871(6) | 1.870(7) |